\documentclass[conference]{IEEEtran}
\IEEEoverridecommandlockouts
\usepackage[font={footnotesize}]{caption}
\usepackage[font={footnotesize}]{subcaption}
\usepackage{amsmath}
\usepackage{svg}
\usepackage{amsfonts} 
\usepackage{booktabs} 
\usepackage{booktabs}
\usepackage{multirow}
\usepackage{enumerate}
\usepackage{xcolor}
\usepackage{graphicx}

\usepackage{url}
\usepackage{pgfplots}
\pgfplotsset{compat=newest} 

\usepackage{graphicx}
\usepackage{tikz}
\usetikzlibrary{external}
\usepackage{scalerel}
\usetikzlibrary{svg.path}

\definecolor{orcidlogocol}{HTML}{A6CE39}
\tikzset{
  orcidlogo/.pic={
    \fill[orcidlogocol] svg{M256,128c0,70.7-57.3,128-128,128C57.3,256,0,198.7,0,128C0,57.3,57.3,0,128,0C198.7,0,256,57.3,256,128z};
    \fill[white] svg{M86.3,186.2H70.9V79.1h15.4v48.4V186.2z}
                 svg{M108.9,79.1h41.6c39.6,0,57,28.3,57,53.6c0,27.5-21.5,53.6-56.8,53.6h-41.8V79.1z M124.3,172.4h24.5c34.9,0,42.9-26.5,42.9-39.7c0-21.5-13.7-39.7-43.7-39.7h-23.7V172.4z}
                 svg{M88.7,56.8c0,5.5-4.5,10.1-10.1,10.1c-5.6,0-10.1-4.6-10.1-10.1c0-5.6,4.5-10.1,10.1-10.1C84.2,46.7,88.7,51.3,88.7,56.8z};
  }
}

\newcommand\orcidicon[1]{\href{https://orcid.org/#1}{\mbox{\scalerel*{
\begin{tikzpicture}[yscale=-1,transform shape]
\pic{orcidlogo};
\end{tikzpicture}
}{|}}}}
\usepackage[hidelinks]{hyperref}

\begin{document}

\title{A Pioneering Roadmap for ML-Driven Algorithmic Advancements in Electrical Networks \vspace{-0.1cm} \\
\thanks{The work of R.J. Bessa and A. Marot were supported through the AI4REALNET project (Grant Agreement No 101119527) and JL Cremer, A. Kelly, and R.J. Bessa are supported through the AI-EFFECT project (Grant Agreement No 101172952); both funded under the European Union's Horizon Europe Research and Innovation program. However, views and opinions expressed are those of the author(s) only and do not necessarily reflect those of the European Union. Neither the European Union nor the granting authority can be held responsible. The authors acknowledge all members of CIGRE C2.42 working group for discussions. \\ {979-8-3503-9042-1/24/\$31.00 © 2024 IEEE} }
}

\author{\IEEEauthorblockN{Jochen L. Cremer}
\IEEEauthorblockA{\textit{Delft University of Technology}, Netherlands \\ \textit{Austrian Institute of Technology}, Austria \\
\small{j.l.cremer@tudelft.nl}}
\and
\IEEEauthorblockN{Adrian Kelly}
\IEEEauthorblockA{\textit{EPRI} \\
Ireland \\
\small{akelly@epri.com}}
\and
\IEEEauthorblockN{Ricardo J. Bessa}
\IEEEauthorblockA{\textit{INESC TEC} \\
Portugal \\
\small{ricardo.j.bessa@inesctec.pt}}
\and
\IEEEauthorblockN{Milos Subasic}
\IEEEauthorblockA{\textit{Hitachi Energy} \\
Germany \\ 
\small{milos.subasic@hitachienergy.com}}
\and
\IEEEauthorblockN{Panagiotis N. Papadopoulos}
\IEEEauthorblockA{\textit{The University of Manchester} \\
United Kingdom \\
\small{panagiotis.papadopoulos@manchester.ac.uk}}
\and
\IEEEauthorblockN{Samuel Young}
\IEEEauthorblockA{\textit{Energy Systems Catapult} \\
United Kingdom \\
\small{samuel.young@es.catapult.org.uk}}
\and
\IEEEauthorblockN{Amar Sagar}
\IEEEauthorblockA{\textit{Arizona State University} \\
United States of America \\
\small{amar.sagar@asu.edu}}
\and
\IEEEauthorblockN{Antoine Marot}
\IEEEauthorblockA{\textit{RTE} \\
France \\
\small{antoine.marot@rte-france.com}}
}

\maketitle

\begin{abstract}
Advanced control, operation, and planning tools of electrical networks with ML are not straightforward. 110 experts were surveyed to show where and how ML algorithms could advance. This paper assesses this survey and research environment. Then, it develops an innovation roadmap that helps align our research community with a goal-oriented realisation of the opportunities that AI upholds. This paper finds that the R\&D environment of system operators (and the surrounding research ecosystem) needs adaptation to enable faster developments with AI while maintaining high testing quality and safety. This roadmap serves system operators, academics, and labs advancing next-generation electrical network tools.
\end{abstract}

\begin{IEEEkeywords}
Machine learning, power system operations, control centre, software, innovation
\vspace{-0.3cm}
\end{IEEEkeywords}

\section{Introduction}
Historically, vendors controlled the entire R\&D development for power systems. This comprehensive approach ensured that the vendor had complete control over every stage of innovation and refinement. 
Today, developments of ML-based algorithms are rapid and often founded in open-source environments, unlocking quick innovation.
However, ML-based approaches have not yet been used to the full extent to empower grid evolution~\cite{Groissbock2019}. 

The power system is critical with special testing and experimentation requirements, and there are several barriers. Furthermore, significant differences exist between conventional software development components and the innovative elements introduced by ML. Conventional software development typically adheres to a structured and deterministic approach, while ML presents aspects of learning, adaptation, and complexity~\cite{Angel2022}. ML developments are typically dispersed and occur in a scattered environment driven by initiatives, projects, and various organisations. In the past few decades, the academic literature and research projects investigated several ML methods and algorithms. However, a systematic overview and roadmap of the implementation of ML methods to power systems is lacking. 

This paper investigates this environment by analysing a survey applied to about $110$ experts (52\% system operators and utilities, $22$\% vendors $26$\% academia) who use or develop ML for the operation of the power grid. Subsequently, we put forward a use case implementation roadmap (or journey) for ML-driven algorithm advances in electrical networks. $63$\% of the survey respondents thought that ML was important for the operation of the power system, specifically to support human decisions, demand and forecasting of renewable energy. However, $50$\% of the respondents experienced scattered interest from multiple departments with isolated prototyping and no common roadmap. 

In response, sensing the urgency, this paper develops a common implementation roadmap for innovating with AI for control centres based on the Technology Readiness Levels (TRLs) for ML systems \cite{MTRL23}. This novel roadmap aims at understanding the challenges, maximising the opportunities of the ecosystem (in particular, for system operators and vendors) and identifying their gaps. Hence, this work calls operators to the role of 'gluing' together the dispersed development environment, ultimately enabling these rapid developments. This paper develops the ML-innovation roadmap for the power system ecosystem for the first time and discusses important requirements and development procedures. 


\vspace{-0.25em}
\section{Innovating with ML}\label{sec:innovation}
The scope of innovation with ML considers different learning paradigms (supervised, unsupervised, and reinforcement learning), their data processing, and the programming code that trains ML models from data sets. The innovation target is a software product for decision support in future control centres \cite{marot2022perspectives} or for autonomous power systems \cite{Sub20}. 

\vspace{-0.25em}
\subsection{Survey findings}

\begin{table}
\centering
\caption{Overview of survey responses regarding low/high priority use cases.}
\label{tab:priority}
\begin{tabular}{l|ccc}
\toprule
\multirow{2}{*}{Use case} & Not relevant/ & Medium/high & Already \\
& low priority [\%] & priority [\%] &in use [\%] \\
\midrule
Forecast load \& DER &	4&	48&	48 \\
Risk assessment&	26	&65	&9\\
Grid monitoring	&27&	65&	8\\
Operations processes&	34&	60&	6 \\
Simulation&	36	&57	&7\\
Market management	&36	&57&	7\\
Unplanned emergency&	35&	60&	5\\
Reporting assistance&	45&	49&	6 \\
\bottomrule\end{tabular}\vspace{-0.6cm}\label{table:operators_ratings}
\end{table}

Software tools in control rooms have requirements specific to each use case and additional requirements when involving ML-based algorithms. Currently, 
around $60$\% of operators expect ML can improve human decisions and forecasting, and $40$\% expect ML to lead to some improvements. ML is already used on around $50$\% of innovations for forecasting demand and renewable energy. The use cases for outage management, planning, the impact of weather events, and frequency risk and uncertainty management are high priority but not currently implemented (about $35$\% of the respondents).

Table~\ref{table:operators_ratings} overviews how the survey respondents rated use cases as high or low priority. The high-priority use cases are promising, where some already implemented ML for forecasting. For example, demand and renewable energy forecasting are high-TRL and commercially available. However, as shown in an analysis of forecasts~\cite{Kazmi2022}, improvements require researching meta, multi-task learning, and spatial-temporal models of forecast errors. Beyond forecasting, operators seem to rank use cases as a high priority that aims at improving situational awareness (risk assessment, monitoring and emergency management). In these cases, AI can outperform human capabilities. Some grid situations require fast assessments and decisions, where humans are too slow. 

However, software involving ML has high safety requirements to avoid side effects, to ensure reliability, and guarantees. The software may have to explain predictions (white/grey box) so operators can build trust \cite{Mar22}. Operators interact with these tools, and learning occurs bi-directional. The software (and ML) has to consider that humans have individual characteristics. People's attitudes towards ML can change drastically and follow the hype cycle. Introducing ML-based software may be perceived as a risk unless the tool and developments consider these drastic fluctuations in perceptions. 

\vspace{-0.25em}
\subsection{Innovation environment}
Innovating with ML shows characteristics that differ from other typical environments. Past development environments were closed laboratories with small expert circles. However, innovating with ML is often a collective effort strongly tied to the open-source environment, e.g., Linux Foundation Energy (LFE)~\cite{Klimt2023}. Versioning and professional coding practices are key for successful collaboration. However, challenges involve versioning, governance, alignment with functional needs, modularity and extension mechanisms (lifecycle), responsibilities, maintenance, and different views on releasing code\cite{Omo23}.

Designing an environment to innovate with ML requires upskilling the workforce. A US study concluded that important skills for AI/ML positions are critical thinking, dealing with multiple data sources, data mining, programming, attitude and general communication \cite{Ver21}. That study distinguishes between AI and ML engineers. ML engineers work more technically and require more technical knowledge, experience, and hard skills than AI engineers. These skills refer to working with data mining, programming, statistics, and big data. In that study, AI engineers have a central position within the project or team and require a holistic process understanding and high communication skills for interaction within and outside the project team and management. 
Innovating with ML involves specific challenges around the quality and content of data, the handling of sensitive and privacy data \cite{Sev22}, compliance with regulatory requirements, building ML expertise and solving problems and development \cite{UR20}. Beyond these data-related challenges, developing modular and object-oriented software is challenging, as ML-based modules (e.g., Python packages) are often updated challenging the management of dependencies.

AI innovation labs address some of these challenges, creating synergies between universities, research centres, and companies that share visions in innovating goals, open-sourcing, and intellectual properties \cite{Hei22}. These labs can accelerate the adoption of agile development environments to keep up with the fast pace of ML algorithms. These laboratories bridge multidisciplinary research with and in ML, focussing on the final application from lower to higher TRLs. These agile researchers in labs are involved in local R\&D teams in companies while performing fundamental research (TRL 1-3).

\begin{figure*}
    \begin{center} \vspace{-1em} 
        \includegraphics[width=0.75\textwidth]{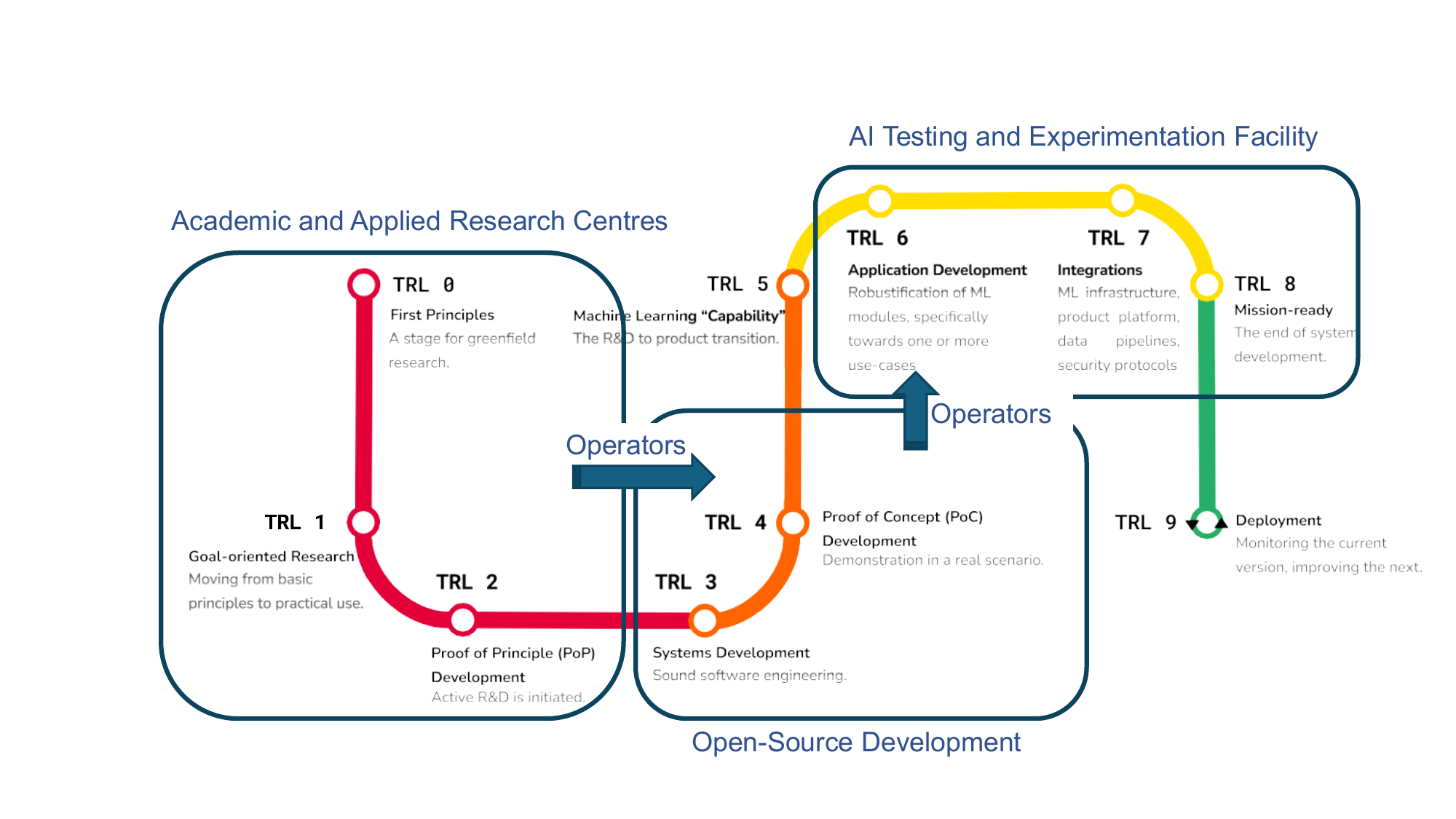} 
    \end{center}
    \caption{Innovation with ML for the electrical network (modified from \cite{MTRL23}). } \vspace{-0.6cm}
    \label{fig:roadmap}
\end{figure*}

\vspace{-0.25em}
\subsection{Interdisciplinary approach}
An interdisciplinary approach explicitly integrates technological, regulatory, and social dimensions during development, innovation, and evaluation. Integrating traditionally distant disciplines, such as computer science and social sciences, ensures a holistic understanding of infrastructures traditionally operated by humans. Although there are a variety of concepts and methods that relate to explainability and trust, there are still many research gaps regarding human-AI-system integration (communication and coordination, social intelligence, human motivation to engage, etc.) or situational awareness in multi-domain operations (e.g., distributed and shared cognition, human expertise, and tacit knowledge). For instance, the Horizon Europe AI4REALNET project approaches interdisciplinarily 
the requirements of operators on the human-AI team, explainability, working conditions, and user-centred and ecological design concepts.


\vspace{-0.25em}
\section{A Use Cases Implementation Roadmap}\label{sec:roadmap}
Fig. \ref{fig:roadmap} shows our roadmap to innovate with ML in this environment. The roadmap involves three 'entities', here simplified and presented in a unidirectional way from the lower to the higher TRL spanning academic and applied research centres, open source development, and AI testing and experimentation (TEF) facilities. A TEF is defined by the Digital Europe Programme as a~\textit{``combination of physical and virtual facilities, in which technology providers can receive primarily technical support to test their latest AI-based software and hardware technologies (including AI-powered robotics) in real-world environments''}. 

\vspace{-0.25em}
\subsection{Developing Proof of Principals (TRL 1-2)}
Research centres typically start with goal-oriented research towards proof-of-principles, initiating research that includes system development. Once principles are developed, individual vendors and system operators (with stronger R\&D departments) scout these principles with their AI/ML experts to investigate creating business value through innovating using these principles. There, around $40$\% of operators already have AI/ML experts, and around $30$\% have an internal AI/ML group. 
$54$\% were involved in processes to identify use cases for ML and develop business cases for organisational adoption. However, often, further development requires collective efforts to share the risk in product development. The means are open-source projects or joined consortiums to jointly innovate. $30$\% of survey respondents would support such collective efforts by sharing data, while half of these require anonymisation. 

The recently published EU AI Act mandates that AI systems providers adhere to a set of ethical principles rather than rigid rules. This underscores the necessity for an ethical-by-design approach, integrating ethical considerations into the developmental framework~\cite{Heymann2023}. Such an approach facilitates the early identification and close monitoring of ethical concerns throughout the research development.

 \begin{figure*}
\centering
\begin{subfigure}{0.18\textwidth} 
        \includegraphics[width=\textwidth]{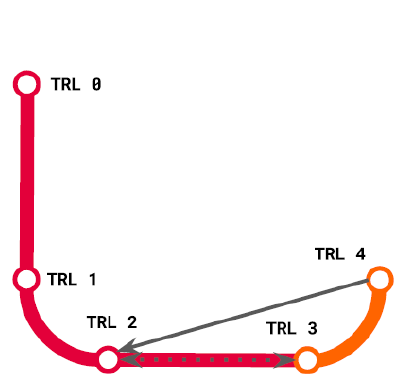}
   \caption{ }
   \label{fig:switchbacka}
\end{subfigure} \hspace{0.04\textwidth}
\begin{subfigure}{0.18\textwidth}
        \includegraphics[width=\textwidth]{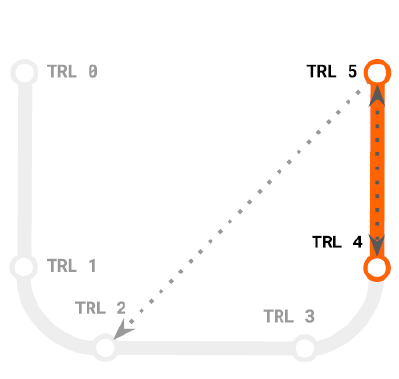}
   \caption{}
\label{fig:switchbackb}
\end{subfigure} \hspace{0.04\textwidth}
\begin{subfigure}{0.21\textwidth} 
        \includegraphics[width=\textwidth]{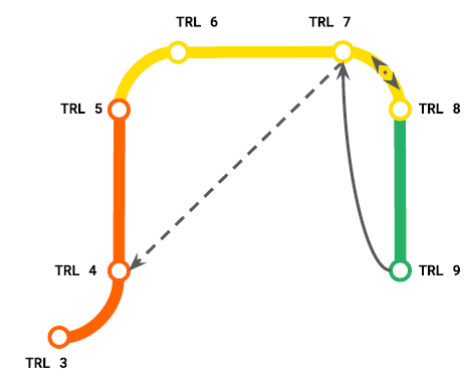}
   \caption{ }
   \label{fig:switchbackc}
\end{subfigure} \hspace{0.04\textwidth}
\begin{subfigure}{0.21\textwidth}
        \includegraphics[width=\textwidth]{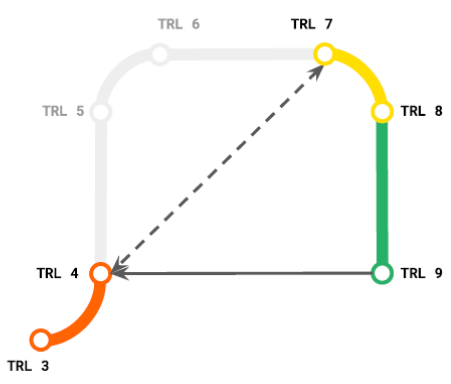}
   \caption{}
\label{fig:switchbackd}
\end{subfigure}
    \caption{Circumstantial discovery switchback (a), predefined, embedded switchback (b). AI starts often at TRL $4$ making it challenging to switchback to lower TRLs. (c) Switchbacks to ML toolboxes if use-case-specific tests and requirements were not met, (d) direct switchback and rapid prototyping between modifying ML toolboxes and final products. Figure from \cite{MTRL23}.}
    \label{fig:switchbacks} \vspace{-1.5em}
\end{figure*}

\vspace{-0.25em}
\subsection{Developing Proof of Concepts (TRL 3-5)}
A consortium or vendor develops a closed or open-source solution from proven principles to proof-of-concept to demonstrate capabilities. Innovating with AI can involve developing concepts further from supervision to self-automation and developing software planning with human-in-the-loop.


These innovation projects are sometimes publicly co-funded research projects (e.g., EU innovation actions or nationally). The LFE or Collaborative Research for Energy System Modelling (CRESYM) associations synthesise the collective efforts of open-source developments. For example, the open-source project Power System Blocks (PowSyBl) is to analyse and simulate grids. 
The governance of the PowSyBl project involves interactions between RTE, the CorNet project, and LFE. Their private code repositories are developed by two parties and contributed to the main library PowSyBl. In such similar projects, around $70$\% of operators took part in three or more projects involving AI, and between $25$-$50$\% of operators use already open-source solutions. 
However, such projects typically face similar challenges. ~$40$\% found it difficult to deploy and integrate with legacy systems, and ~$40$\% had difficulties in standardising data access and interfaces. For example, multiple incomplete copies of data may exist. The design of testbeds, controls, verification, and certification methods for AI-based algorithms can provide the requirements for standardisation and interfacing different tools and data.

\vspace{-0.1cm}
\subsection{Testing and Experimentation (TRL 6-9)}
Subsequently, the owner of the product is interested in certifying, testing, and experimenting with the software developed to make it ready for deployment. As the electricity grid is a critical infrastructure and ML technology has specific characteristics different from other technologies, a novel way to experiment and test the technology is required. This means experimenting, testing, and certifying AI-based tools deployed to distribution and transmission system operators, and also considering the secure interface between distributed energy resources and the grid. 
These tests and experimentation are urgently needed to integrate the software tool with the existing control centre as $65$\% of respondents found low-quality data sets or a general lack of mappings to link multiple sources as a barrier. 
Overall, there are two types of facilities: internal and shared facilities. The internal facilities of a system operator verify the functionalities of the developed tool and make final improvements before deployments. In shared virtual and physical facilities such as TEF, testing and experimentation costs can be shared between system operators.

ML testing differs from testing classical software~\cite{Zhang2022}. While software testing primarily identifies bugs within the code itself, ML testing focuses on rectifying errors within input data, the learning process, structural aspects, and the model's hyperparameters. The response of an ML system can evolve with updates to input and historical data, contrasting with the generally static nature of software code. The forms of test input vary in ML, encompassing input data for training and operation, as well as the functions of the model. Additionally, ML testing often yields more false positives in bug detection compared to traditional software testing. Bugs in ML systems may manifest in data, code, or the mathematical algorithms employed, necessitating proactive involvement from ML engineers throughout the testing phase. Testing should also capture non-functional requirements such as fairness, data privacy protection, robustness and security, explainability, and interoperability.

\subsection{Project Management}
ML project management has technical, organisational, and ecosystem challenges that can trigger changes. 
Technical and organisational challenges might relate to
\begin{itemize}
\item Data (accessibility, availability, quality), models (performance, interpretability, scalability), and experimentation (slow iteration, reproducibility issues).
\item compute and infrastructure, inefficient technical stack (not evolutive, not interoperable, mismatch between development and production environments), keeping up to date with rapidly evolving software versions, costs
\item not well-defined business use case, lack of funding and manager-buy-in, lack of skills and roles for deployment, end-user requirements, and adoption, regulation, and compliance, managing errors
\end{itemize}

Furthermore, integrating automated machine learning (AutoML) methods requires a re-evaluation of the organisational structure and workflows within ML teams, alongside the imperative to foster proficiency with these emerging software solutions. As illustrated in~\cite{Kreuzberger2023}, Machine Learning Operations (MLOps) 
engineers assume a multifaceted role that combines expertise from various domains, including data science, data engineering, software engineering, DevOps, and backend development. This interdisciplinary role encompasses the establishment and management of the ML infrastructure, the orchestration of automated ML workflow pipelines, the facilitation of model deployment into production environments, and the ongoing monitoring of models and ML infrastructure.

\vspace{-0.25em}
\section{Switchbacks}\label{sec:Switchbacks}
The roadmap is not unidirectional as Fig. \ref{fig:switchbacks} shows.

\vspace{-0.5em}
\subsection{Circumstantial Discovery Switchback}
A project could switch back from TRL $4$ to $2$ (Fig. \ref{fig:switchbacka}) if the chosen model or problem formulation does not produce satisfactory performance. This switchback is typically triggered when new technical gaps emerge during system integration, prompting multiple rounds of development iterations. For example, one might need to better deal with unstructured or continuous data types or additional features. One might need to consider larger models or models with more appropriate architecture to leverage some domain knowledge and problem structure. For example, physical constraints could be used to support learning the model~\cite{Misyris2020}. One might also reconsider the problem formulation in terms of target predictions or learning objectives. For instance, one could switch from regression to classification if qualitative results are sufficient, such as classifying grid states without predicting all the power flows.

\subsection{Predefined, Embedded Switchback}
Switching between TRL $4$ and $5$ (Fig. \ref{fig:switchbackb}) is common in ML due to its experimental nature, requiring iterative experiments for performance improvement. Sandbox environments should be first created to allow for quick experiments and updates. Ideally, experiments should first be run in controlled environments (e.g., Grid2Op\footnote{\vspace{-2em} https://github.com/rte-france/Grid2Op}) with known ground truth using clean synthetically generated data. In this way, one could distinguish between issues in problem formulation and model selection. In the second stage, one can identify issues related to real data quality (missing, noisy, incomplete). Using real data directly may lead to unsatisfactory performance and might switch us back to TRL $2$ to rethink the model or problem formulation. However, it could just be a matter of improving data quality. Establishing a targeted evaluation scheme helps to identify performance hindrances and enables quick iterations. This evaluation scheme set-up should be ideally decoupled from the model development and be model-agnostic to avoid overfitting performance on a given solution.

Switching back from TRL $5$ to $2$  (Fig. \ref{fig:switchbackb}) could be needed after evaluating the performance in specific examples, possibly with the help of end users. The nonuniform accuracy may become a concern, especially with a lack of interpretability or uncertainty estimation. Poor code usability, configuration complexity, and unclear workflows can require substantial revisions at the initial design stage. Simplicity, reproducibility, and reusability are crucial for ML to avoid accumulating technical debt during deployment, maintenance, and future iterations at TRL $5$-$6$.  Inadequate development tools, experiment versioning, workflow structuring, and data history can impede project management. Establishing an end-to-end workflow with simple models to gauge baseline performance supports identifying areas that need more attention and preventing overemphasis on specific modules that may not improve overall performance. 

\subsection{Switchback When Requirements Not Met}
Advancing to a higher TRL for production requires stricter adherence to recent data, internal IT systems, and external regulations. Considering continuous end-user feedback via model retraining may prompt a switchback from $9$ to $7$ (Fig. \ref{fig:switchbackc}). One might also want to log continuous feedback and preferences from end-users and align the application through model retraining, triggering a switchback. Considerations for model performance monitoring, retraining cycles, and traceability are crucial in the AI lifecycle, necessitating extensive testing with end-user datasets, e.g., switching from $8$ to $7$.

A review switchback from TRL $7$ to $4$ may be triggered at review gates. Such 'gates' are key decision points in the project lifecycle where specific components of the ML system may require further development and are sent back to an earlier TRL. These may mitigate performance drops due to a mismatch between the development and production environment, which leads to a decrease in the expected performance. 
Bridging the gap between these environments, especially when reliant on open-source versus proprietary technologies, may delay progression until the deployment stack is updated. 

Open sourcing often requires different APIs and data transformations for broad usability, which remains crucial later. The exchange of data pipeline considerations at both levels $4$-$5$ and $7$-$8$ are often fruitful for both sides. When improving the model, for example, by modifying data features, cycling back to level $7$ can test for unintended consequences or biases. Similarly, functional features, such as active learning, might integrate well, but still require some development. One issue is often the availability of historical data. Augmenting the data set with partially synthetic data could help. However, caution is needed with models pre-trained on simulated data and fine-tuned on real data to avoid miscalibration. Switching back from level $7$ to $4$ can address this. 

\subsection{Direct, Planned Switchbacks}
These are also planned switchbacks in the ML TRL process (see Fig. \ref{fig:switchbackd}) that are not encouraged \cite{MTRL23} as they reduce development speeds. Such planned switchbacks range from $9$ to $4$ (or $9$ to $2$) and are often intended to consider field feedback and incorporate advancements in research. 

\section{Conclusions}\label{sec:conclusions}

Urgent innovation is needed for electrical networks. Our survey highlights opportunities and challenges when innovations use ML. Innovating electrical infrastructure requires extensive testing, experimentation, and regulations that require special ethical considerations. Realising faster development of ML-based solutions, as already realised in other domains, requires attention to a changing research and development environment, such as enabling AI innovation laboratories and open datasets. In response, this paper presents an innovation roadmap (journey of AI) to collectively realise new use cases and enhance existing ones.  

\bibliographystyle{IEEEtran}
\bibliography{ref.bib}
\end{document}